# Hierarchical Control of Grid-Connected Hydrogen Electrolyzer Providing Grid Services

Bang L.H. Nguyen[†§], Mayank Panwar[§], Rob Hovsapian[§], Yashodhan Agalgaokar[§], Tuyen Vu[†]
[†]*ECE Department, Clarkson University*, Potsdam, NY, USA
[§]*National Renewable Energy Research Laboratory*, Golden, CO, USA
bang.nguyen@nrel.gov, mayank.panwar@nrel.gov, rob.hovsapian@nrel.gov, yagalgao@nrel.gov, tvu@clarkson.edu

*Abstract*— This paper presents the operation modes and control architecture of the grid-connected hydrogen electrolyzer systems for the provision of frequency and voltage supports. The analysis is focused on the primary and secondary loops in the hierarchical control scheme. At the power converter inner control loop, the voltage- and current-control modes are analyzed. At the primary level, the droop and opposite droop control strategies to provide voltage and frequency support are described. Coordination between primary control and secondary, tertiary reserves is discussed. The case studies and real-time simulation results are provided using Typhoon HIL to back the theoretical investigation.

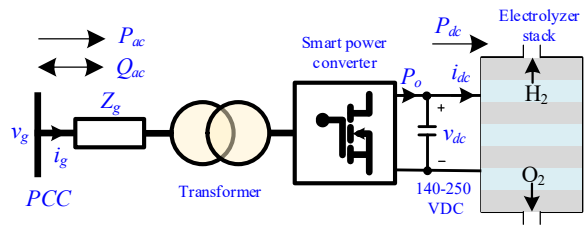

*Figure 1. Circuit diagram of a grid-connected electrolyzer system.*

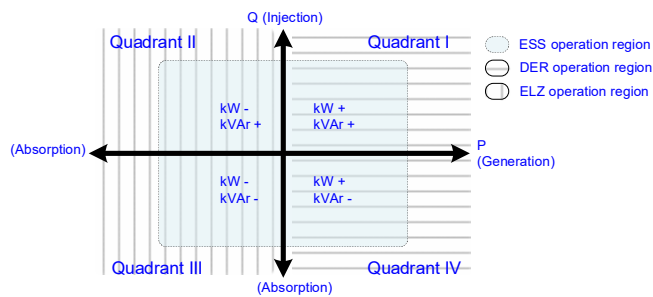

*Figure 2. Comparison of the operation areas between ESS, DER, and electrolyzer (ELZ) in power flow quadrant diagram.*

## I. INTRODUCTION

Hydrogen electrolyzers are energy conversion systems for a hydrogen-based energy storage system (ESS) that can absorb excess energy from the intermittent generation of renewable energy resources. Compared to other ESS such as battery and flywheels, electrolyzer has a low environmental impact, high energy density, and has no energy leakage [1]. The excessive energies turned into hydrogen that can be used later in other industrial processes, transportation, and power generation.

Hydrogen electrolyzer can be employed to reduce the impact of variability on load fluctuations and stability of the electric grid with renewable and distributed energy resources (DER) [2]. Several papers investigated the potential of electrolyzers in grid supports. In [3], the authors outline the dynamic characteristics of an electrolyzer and its ability in providing grid services including end-user energy management, supports for transmission and distribution systems, integration of renewables, and wholesale electric market services. The work in [4] proposes a generic front-end control for electrolyzer to enhance the grid flexibility. A 25 MW PEM electrolyzer is installed in Belgium to provide grid services such as grid balance and frequency containment reserve [5]. In Nether lands, a 1-MW pilot electrolyzer is installed to investigate the potential of frequency support [6]. The authors in [7] utilize the hydrogen energy system as a grid management tool to stabilize the grid frequency in the Hawaiian island grid. In [8], it is proved that the dispatchable electrolyzer can reduce both the frequency variation and the settling times under load changes, loss of generation, and line faults. The authors in [9] prove that electrolyzer fleet can benefit the distribution voltage profiles under the impact of solar-photovoltaic-based generation.

Fig. 1 shows the circuit diagram of an electrolyzer system including a smart power converter, transformer, and the connection line interfacing the grid at the point of common coupling (PCC). Both the active and reactive power $P_{ac}$, $Q_{ac}$ of the electrolyzer system can be regulated to support the grid operation. Notably, $P_{AC}$ can only be consumed, while $Q_{ac}$ can be both injected and absorbed by the electrolyzer system. The consumption power $P_{dc}$ of the electrolyzer stack equals the active power $P_{ac}$ subtracted the loss. Fig. 2 shows a 4-quadrant diagram, where the operation region of electrolyzer (ELZ) is compared with those of generation resources (DER) and ESS.

From the point of view of the demand-supply balance, instead of increasing power generation, one can decrease the electrolyzer power and vice versa. The electrolyzer can reduce its power to support the demand rise or consume more power in case of the load power suddenly drops.

In most of the existing studies, electrolyzers are mainly adopted as demand response resources with only active power regulation ability, whereas the voltage support by regulating the reactive power is omitted. Moreover, the operation modes of the grid-supporting smart power converters [10] are ignored except for [11], where power-electronics interfaces and controls are investigated. However, [11] mainly focused on the fast frequency response with the proposed active power-frequency curve. In this paper, we study the operation modes and control architecture of the grid-supporting power converter for the hydrogen electrolyzer providing both voltage and frequency supports.

The paper is organized as follows. In Section II, the operation modes of the power converter at the inner control loop are presented; and the droop and opposite controls for each operation mode are developed and presented. In Section III, we study the secondary control and overview the control architecture of the grid-connected power converters for grid supports. The electromagnetic transient (EMT) real-time simulations using Typhoon HIL for an electrolyzer system supporting a DER are presented in Section IV. Conclusion and future work are presented in Section V.

## II. OPERATION MODES AND PRIMARY DROOP CONTROLS

The power converter for an electrolyzer flexible load can be controlled in two operation modes: the *voltage control mode and current control mode* as shown in Fig. 3. These operation modes are similar to those of generation resources [12] except that the current direction is reversed.

### A. Voltage-Control -Mode and Droop Rules

Fig. 3(a) shows the block diagram of this mode, where the inner loop of the power converter regulates the output voltage and frequency following references generated by the primary control loop. The primary control can simply determine the voltage $E^*$ and frequency $f^*$ references based on droop rules as follows.

$$f^* = f^‡ + k_p(P_{AC} - P^‡) \quad (1)$$

$$E^* = E^‡ + k_q(Q_{AC} - Q^‡) \quad (2)$$

where $k_p$ and $k_q$ are the droop coefficients; $f^‡$ is the nominal frequency; $E^‡$ is the nominal voltage magnitude; $P^‡$ and $Q^‡$ are the active and reactive power set-points at the nominal frequency and voltage, respectively; $P_{AC}$ and $Q_{AC}$ are the active and reactive power measurements. Other control functions can be included in the primary control to fulfill different objectives or improve the control performances [13]. Fig. 4 shows the electrolyzer power regulation characteristics under the droop rules.

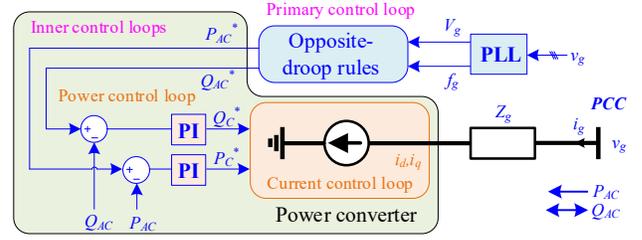

Figure 3. Control diagram and equivalent circuit of a power converter operating in (a) voltage, and (b) current control modes.

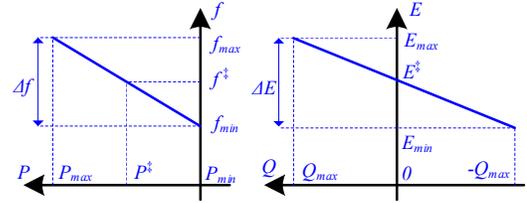

Figure 4. Electrolyzer power regulation characteristics with droop control.

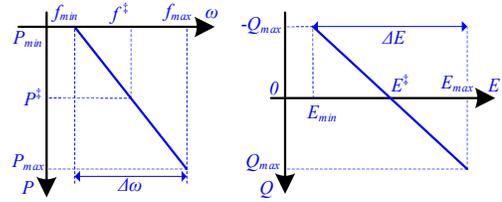

Figure 5. Voltage and frequency regulation characteristics with opposite droop control.

### B. Current-Control-Mode and Opposite Droop Rules

Fig. 3(b) shows the block diagram of this mode, where the inner loop of the power converter regulates the output currents following references generated by the primary control loop. The primary control performs the current reference computation based on the opposite droop rules [14] as follows. The incoming power from the grid to the power converter system can be calculated as

$$P_{AC} = v_d i_d + v_q i_q, \quad (3)$$

$$Q_{AC} = v_q i_d - v_d i_q. \quad (4)$$

Form (3) and (4), the reference currents can be obtained from the reference powers as follows:

$$i_d^* = \frac{v_g^d P_{AC}^* + v_g^q Q_{AC}^*}{\sqrt{\left(v_g^d\right)^2 + \left(v_g^q\right)^2}} \quad (5)$$

$$i_q^* = \frac{v_g^q P_{AC}^* - v_g^d Q_{AC}^*}{\sqrt{(v_g^d)^2 + (v_g^q)^2}} \quad (6)$$

The power references can be obtained from the opposite droop control as

$$P_{AC}^* = P^{\ddagger} + k_f(f - f^{\ddagger}) \quad (9)$$

$$Q_{AC}^* = Q^{\ddagger} + k_v(E - E^{\ddagger}) \quad (10)$$

Where $k_f$ and $k_v$ are the reverse droop coefficients, and $E$ & $f$ are the measured magnitude and frequency of the PCC voltage. The opposite droop mechanism compensates for the voltage and frequency deviation by adjusting the active and reactive powers. The voltage and frequency regulation characteristics posing by the opposite droop control are shown in Fig. 5.

## III. ELECTROLYZER CONTROL ARCHITECTURE

### A. Secondary Controls

As can be seen in equations (1) and (2), the primary droop rules adjust the output voltage magnitude and frequency proportionally to the differences between the output power and the power setpoints. The secondary control is responsible for directing the voltage frequency back to its nominal value. The secondary regulation terms ($\Delta f, \Delta E$) are added into the primary droop controls as follows.

$$f^* = f^{\ddagger} + k_p(P_{AC} - P^{\ddagger}) + \Delta f \quad (11)$$

$$E^* = E^{\ddagger} + k_q(Q_{AC} - Q^{\ddagger}) + \Delta E \quad (12)$$

The PI controller can be employed here to generate the secondary regulation terms as follows.

$$\Delta f = K_{PF}(f^{\ddagger} - f) + K_{IF}\int(f^{\ddagger} - f)dt, \quad (13)$$

$$\Delta E = K_{PV}(E^{\ddagger} - \bar{E}) + K_{IV}\int(E^{\ddagger} - E)dt, \quad (14)$$

where $K_{PF}$ and $K_{IF}$ are the proportional and integral gains in the PI controller for frequency; $K_{PV}$ and $K_{IV}$ are the proportional and integral gains in the PI controller for voltage magnitude; $\bar{E}$ is the average magnitude of the bus voltages.

In the opposite droop rules, when the secondary control is running, the frequency equals its nominal values. The active power correspondingly equals to its setpoint. Therefore, the secondary regulation terms are added to compensate for this effect.

$$P_{AC}^* = P^{\ddagger} + k_f(f - f^{\ddagger} + \Delta f) \quad (15)$$

$$Q_{AC}^* = Q^{\ddagger} + k_v(E - E^{\ddagger} + \Delta E) \quad (16)$$

### B. Design of Power Setpoints

The setpoints of active and reactive powers are significant to the response of the electrolyzer. Without secondary regulation, the electrolyzer power equals its setpoints at the nominal frequency and voltage magnitude. When the frequency and voltage magnitude deviates from their nominal values, the electrolyzer adjusts its powers to support. At nominal frequency, the electrolyzer system have the margins of the active power quantity of $(P^{\ddagger} - P_{min})$ and $(P_{max} - P^{\ddagger})$, where $P_{max}$ and $P_{min}$ are the upper and lower limits of the consumption power of the electrolyzer system. Similarly, the electrolyzer system have the reactive power margins of $(Q^{\ddagger} - Q_{min})$ and $(Q_{max} - Q^{\ddagger})$ at the nominal voltage.

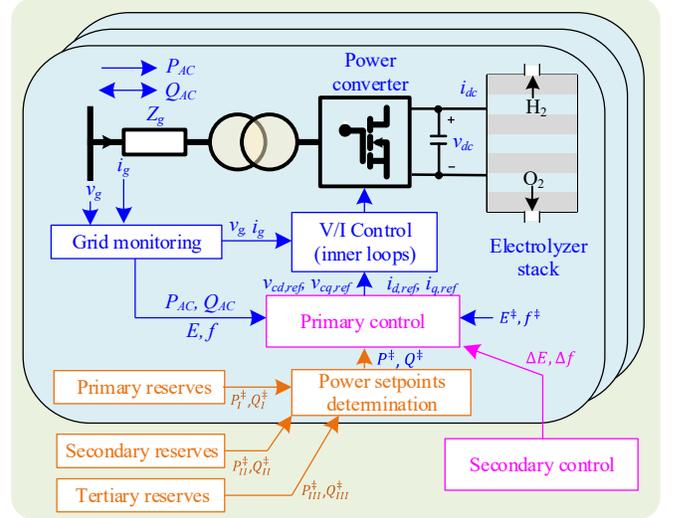

Figure 6. Block diagram of the control architecture of grid-connected hydrogen electrolyzers providing voltage and frequency supports.

Hence, the design of power setpoints determines the power reserves of the electrolyzer to support the system in case of demand rises or drops. On one hand, the electrolyzer owner wants to maximize the active power consumption to achieve the highest hydrogen production and minimize the reactive one to reduce the switching loss. On the other hand, we need the power reserves to support the voltage and frequency. Therefore, the active and reactive power setpoints can be determined by solving an optimization problem. The formulation of this optimization problem is the future work.

### C. Electrolyzer Control Architecture

The control architecture of the grid-connected hydrogen electrolyzer provides voltage and frequency supports are shown in Fig. 6. The analysis of each functional block is given as follows. The grid monitoring block represents the measurement process to collect the grid voltage $v_g$, the incoming current $i_g$. From these measurements, the incoming powers $P_{AC}$ and $Q_{AC}$, the frequency $f$ and voltage $E$ magnitude can be calculated.

Representing by the block *V/I control*, the inner loops regulate the power dispatch following the voltage references or the current references, which are provided from the primary control. There, voltage (V) or current (I) control modes can be implemented respectively as in Section II. To guarantee the control performance, these inner loops with the close-loop controls with feedback signals should be designed [15].

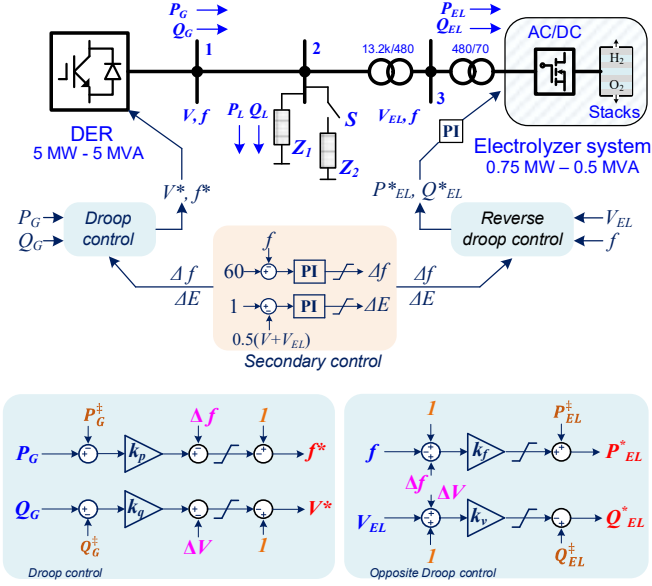

Figure 7. Hydrogen electrolyzer supporting a microgrid with a DER.

The *primary control* generates the voltage ($v_{cdref}, v_{cqref}$) or current references ($i_{dref}, i_{qref}$) to the inner control using the droop or opposite droop rules. Other controls for autonomous responses such as inertia service, the fast frequency response can also be implemented here [16] to provide better response. The nominal values of voltage magnitude and frequency are $E^{\ddagger}$ and $f^{\ddagger}$, respectively.

The power setpoints ($P^{\ddagger}$, $Q^{\ddagger}$) are determined by coordinating the reserves at primary, secondary, and tertiary levels. The *primary reserves* ($P_I^{\ddagger}, Q_I^{\ddagger}$) are responsible for power margins to provide autonomous services [16]. The *secondary reserves* ($P_{II}^{\ddagger}, Q_{II}^{\ddagger}$) are employed to compensate for the steady-state errors in the grid voltage and frequency. The *tertiary reserves* ($P_{III}^{\ddagger}, Q_{III}^{\ddagger}$) are responsible for economically optimizing the grid operation and managing eventual contingency. In this paper, we assume that the power setpoints are already determined.

The secondary control collects the frequency and average voltage throughout the system. Then, a centralized controller is employed to generate the secondary regulation terms ($\Delta E, \Delta f$) which direct the voltage frequency back to its nominal value.

## IV. SIMULATION RESULTS

The study cases are carried out with a DER connecting with loads and an electrolyzer system as shown in Fig. 7. The DER with 5 MW and 5 MVA rated powers are regulated by the droop control. There are two *RL* loads connected at bus 2; $Z_1$ is a fixed load while $Z_2$ is linked via the breaker *S*. The electrolyzer stack is connected at bus 3 via a 480 V/70 V transformer and an AC/DC power converter. There is a 13.2 kV/480 V transformer connected before bus 3. The rated powers of the electrolyzer system are 0.75 MW and 0.5 MVA. The power dispatch of the electrolyzer is regulated via the opposite droop control at the outer loop and PI controller at the inner loop. The secondary control is implemented with two PI controller which regulate the frequency toward the nominal value of 60 Hz and the average voltage of buses 1 and 3 toward the nominal value of 1 per unit. The detail of control parameters is provided in Appendix. The load change events are made by closing *S* at 2.2 (s) and opening *S* at 3.2 (s) i.e., the load $Z_2$ is connected and disconnected, respectively. The study cases compare the responses in the DER with and without electrolyzer support in cases with and without secondary control. In the case without supports from the electrolyzer, the electrolyzer powers are kept constantly at 400 kW and -100 kVA.

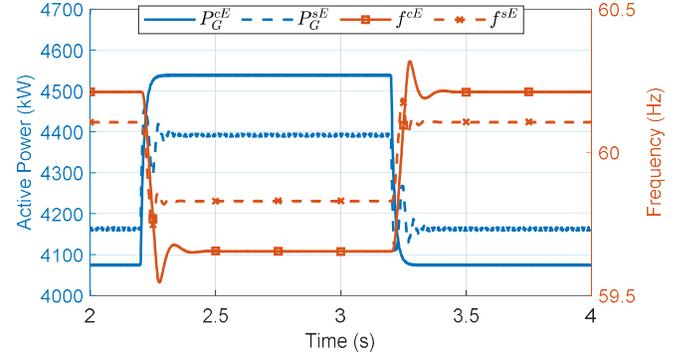

Figure 8. Active power of the DER with constant electrolyzer ($P_G^{cE}$) and supporting electrolyzer ($P_G^{sE}$); frequency response with constant electrolyzer ($f^{cE}$) and supporting electrolyzer ($f^{sE}$) under no secondary control.

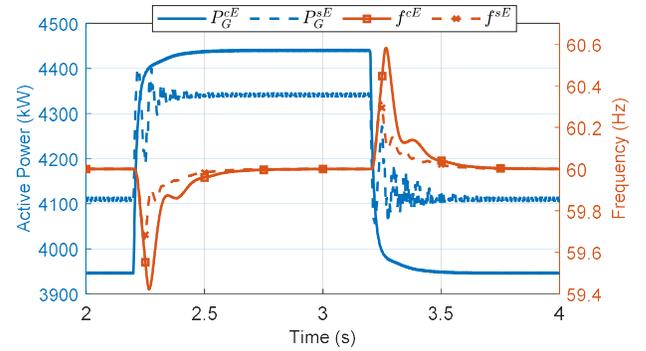

Figure 9. Active power of the DER with constant electrolyzer ($P_G^{cE}$) and supporting electrolyzer ($P_G^{sE}$); frequency response with constant electrolyzer ($f^{cE}$) and supporting electrolyzer ($f^{sE}$) under secondary control.

Fig. 8 compares the DER active powers in the case of the constant-power electrolyzer ($P_G^{cE}$) with the supporting electrolyzer ($P_G^{sE}$) under no secondary control. One can see that, with the constant-power electrolyzer, the DER active power needs to increase from above 4050 kW to above 4500 kW. Whereas, with the supporting electrolyzer, the DER active power increases only from above 4150 kW to 4400 kW. The DER active power variation is reduced from about 450 kW to 250 kW. In the same figure, the frequency deviation is also smaller with the supporting electrolyzer ($f^{sE}$) compared to without electrolyzer support ($f^{cE}$). Fig. 9 shows the same parameters as Fig. 8 but under secondary control. In this case, the frequency deviation occurs under the sudden load change and then is regulated back to its nominal values by the secondary control. Like Fig. 8, the power and frequency variations are reduced with the supporting electrolyzer.

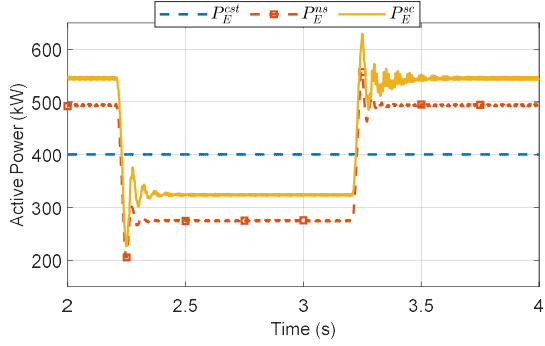

*Figure 10. Active power of electrolyzer: constant power ($P_E^{cst}$), no secondary control ($P_E^{ns}$), and under secondary control ($P_E^{sc}$).*

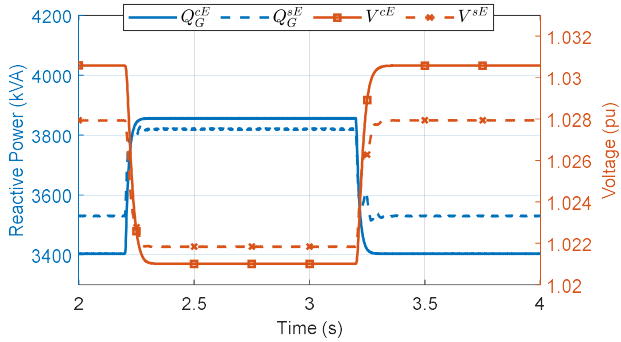

*Figure 11. Reactive power of the DER with constant electrolyzer ($Q_G^{cE}$) and supporting electrolyzer ($Q_G^{sE}$); voltage response with constant electrolyzer ($V^{cE}$) and supporting electrolyzer ($V^{sE}$) under no secondary control.*

Fig. 10 shows the active power response of the hydrogen electrolyzer. The constant power ($P_E^{cst}$) of the electrolyzer is maintained at 400 kW. Under the load change that occurred at 2.2 (s), the active power of the electrolyzer with secondary control is reduced from about 550 kW to near 350 kW. Then, when the load is disconnected, the electrolyzer active power is returned to 550 kW. The active power responses under no secondary control ($P_E^{ns}$) is smaller than those with secondary control ($P_E^{sc}$) about 50 kW with the same response.

Fig. 11 compares the DER reactive powers in the case of the constant-power electrolyzer ($Q_G^{cE}$) with the supporting electrolyzer ($Q_G^{sE}$) under no secondary control. In the same figure, the voltage deviation is also compared between the case with supporting electrolyzer ($V^{sE}$) and with constant-power electrolyzer ($V^{cE}$). Both reactive and voltage deviations are smaller with the supporting electrolyzer. Fig. 12 shows the same parameters as Fig. 11 but under secondary control. The secondary control coordinates the average voltage toward its nominal values. Like Fig. 11, with electrolyzer supports, the reactive power and voltage variations are smaller than those of with constant power electrolyzer.

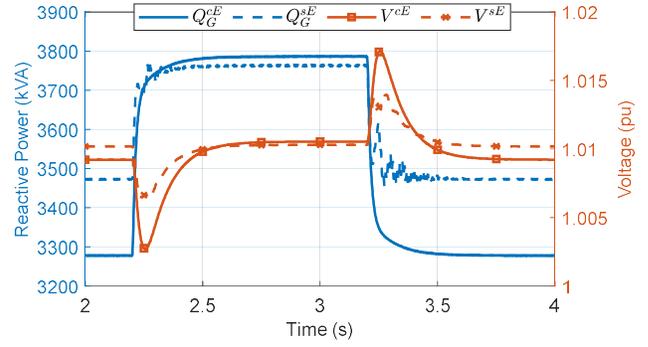

*Figure 12. Reactive power of the DER with constant electrolyzer ($Q_G^{cE}$) and supporting electrolyzer ($Q_G^{sE}$); voltage response with constant electrolyzer ($V^{cE}$) and supporting electrolyzer ($V^{sE}$) under secondary control.*

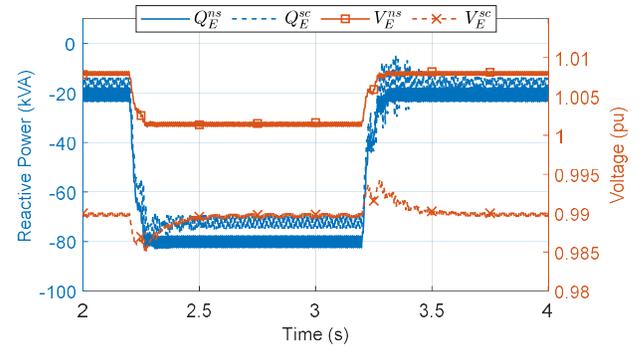

*Figure 13. Reactive power of electrolyzer: no secondary control ($Q_E^{ns}$), and under secondary control ($Q_E^{sc}$); the terminal voltage at the PCC connecting to electrolyzer system: no secondary control ($V_E^{ns}$), and under secondary control ($V_E^{sc}$).*

In Fig. 13, the reactive powers of the electrolyzer system are compared between the cases with and without secondary control. These two cases yield nearly the same reactive powers owing to the secondary regulation terms. In the same figure, the voltage response with secondary control is smaller than those without secondary control.

In summary, one can see that the electrolyzer supports can reduce the voltage magnitude and frequency deviations. By regulating the active and reactive powers reasonably from the electrolyzer side, the needs for the active and reactive power response from DER are mitigated. The secondary control performed well in restoring the frequency to its nominal value. The real-time simulations are carried out using Typhoon HIL.

## V. Conclusion and Future Works

This paper study the operation modes and the control architecture of the grid-connected hydrogen electrolyzer providing voltage and frequency supports. The voltage control mode and current control mode are analyzed. The droop and opposite droop controls for electrolyzer are derived from the traditional ones in DER control. The simulation results prove that an electrolyzer can effectively provide frequency and voltage supports. The grid-supporting controls impact hydrogen production. Therefore, one needs to optimize the electrolyzer operation considering the benefits given by providing the grid supports and other economic criteria.

Future works would provide more detailed analysis and design of electrolyzer fleets for low-inertia grids with better control designs and power-hardware-in-loop implementation. The real-time simulation using Typhoon HIL would be added with the real electrolyzer system hardware and external controller for the grid-connected power converters.

## Acknowledgement

This work was authored in part by the National Renewable Energy Laboratory, operated by Alliance for Sustainable Energy, LLC, for the U.S. Department of Energy (DOE) under Contract No. DE-AC36-08GO28308. Funding provided by U.S. Department of Energy Office of Energy Efficiency and Renewable Energy Hydrogen and Fuel Cell Technologies Office. The views expressed in the article do not necessarily represent the views of the DOE or the U.S. Government. The U.S. Government retains and the publisher, by accepting the article for publication, acknowledges that the U.S. Government retains a nonexclusive, paid-up, irrevocable, worldwide license to publish or reproduce the published form of this work or allow others to do so, for U.S. Government purposes.

## Appendix

*DER Droop Control:* Rated active powers: $P_{G-rated} = 5$ MW, Rated reactive powers: $Q_{G-rat} = 5$ MVA, Active power setpoints: $P_{G-set} = 0.85$ (pu), Reactive power setpoints: $Q_{G-set} = 0.85$ (pu), Nominal line-line RMS voltage: 13.2 kV, Nominal frequency: 60 Hz, Frequency droop gain: 0.1, Voltage droop gain: 0.1, Low-pass filter for active power: $T_f = 0.02$.

*Electrolyzer Reverse Droop Control:* Rated active powers: $P_{EL-rated} = 0.75$ MW, Rated reactive powers: $Q_{EL-rated} = 0.5$ MVA, Active power setpoints: $P_{EL-set} = 0.4$ MW, Reactive power setpoints: $Q_{EL-set} = 0$ MVA, Active power droop gain: 52500, Reactive power droop gain: $10^4$. PI gains (in both acitve and reactive powers): $K_I = 0.1, K_P = 10$.

*Secondary control loop:* PI gains (in both voltage and frequency): $K_I = 0.1, K_P = 10$, Execution rate: 10 *ms*.